\documentclass{cpcauth}

 \usepackage{epsfig}

\usepackage{amssymb}

\begin{document}

\begin{frontmatter}



\title{FOLDING IN LATTICE MODELS
WITH SIDE CHAINS}


\author{Mai Suan Li$^1$, D. K. Klimov$^2$ and D. Thirumalai$^2$}

\address{$^1$Institute of Physics, Polish Academy of Sciences,
Al. Lotnikow 32/46, 02-668 Warsaw, Poland\\
$^2$Department of Chemistry and Biochemistry and Institute for Physical
Science and Technology, University of Maryland, College Park, MD 20742 }

\begin{abstract}
The folding kinetics of three-dimensional
lattice Go models with side chains is studied using two different
Monte Carlo move sets. 
A flexible move set based on  single, double and triple backbone moves
is found to be far superior  compared to the
standard Monte Carlo dynamics. 
In accord with previous theoretical predictions we find
that 
the folding time grows as a power
law with the chain length and the corresponding exponent $\lambda \approx 3.7$
for Go models.
The study shows that the incorporation of side chains dramatically slows
down folding rates. 
\end{abstract}

\begin{keyword}
Protein folding \sep lattice model \sep side chain \sep Monte Carlo
\PACS  71.28.+d, 71.27.+a
\end{keyword}
\end{frontmatter}

In recent years, considerable insight into the thermodynamics and kinetics 
of protein
folding has been gained due to simple lattice and off-lattice models 
\cite{Dill,Thirumalai0} with a small
number of beads. In these toy models of proteins, the beads represent amino
acids. The length  of single domain proteins $N$ ranges approximately 
from 30 to 200. 

The  dependence of folding times, $t_f$, on $N$ 
is an interesting  problem in protein physics. 
Based on analogy to polymer physics Thirumalai \cite{Thirumalai}  
predicted that
the folding times, $t_f$, should grow with the chain length,
$N$, by power laws, i.e 
\begin{equation}
t_{f} \; \; \sim \; \; N^{\lambda} \; 
\label{eqn1}
\end{equation}
if folding proceeds  through direct pathways with a nucleation mechanism.
This prediction has been supported by later studies
\cite{Shakhnovich,Cieplak}.
For simple two-state folders $\lambda$ was estimated to be between 3.8 and 4.2
\cite{Thirumalai}. 
Numerical studies using various lattice models
{\em without side chains} (SC) \cite{Shakhnovich,Cieplak} 
indicated the dependence of $\lambda$ on
specifics of the model, dimensionality and temperature, $T$.
For the optimal temperature $T_{min}$ ($\approx T_F$, the folding
temperature), at which  folding is fastest,
$\lambda \approx 6$ and $\approx 4$
for  random and designed sequences, respectively \cite{Shakhnovich}.
For the Go model \cite{Go} $\lambda \approx 3$
\cite{Shakhnovich,Cieplak}. 
In these studies $t_f$ is defined as a median value of the first passage times.

It is well known that lattice models in which only $\alpha$-carbons are
represented by single beads, are oversimplified models of 
real proteins. A next natural step to mimic more realistic features of
proteins  such as a dense core packing 
\cite{Klimov}
is to include the rotamer degrees of freedom.
One of
the simplest models is a cubic lattice of a backbone (BB) sequence 
of $N$ beads,
to which a side bead representating a SC is attached
\cite{Klimov}. The system has in total 2$N$ beads.
Fig. 1 shows  a typical native conformation with a SC for $N=18$.
The kinetics and thermodynamics of sequences with SCs were studied
by Klimov and Thirumalai \cite{Klimov}. The SCs
were shown \cite{Klimov} to enhance the  
cooperativity of folding. 

\begin{figure}
\epsfxsize=2.5in
\centerline{\epsffile{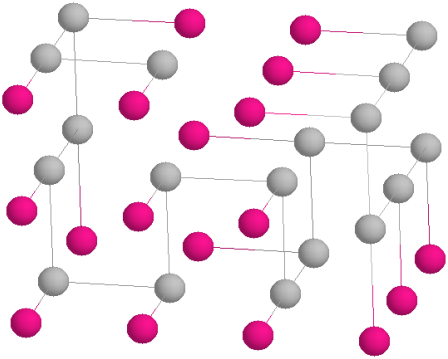}}
\caption{Native conformation of lattice sequence with SC ($N=18)$. 
The BB and SC beads are denoted by grey and black circles,
respectively.}
\end{figure}

The aim of this paper is twofold. First,
we try to compare the efficiency of the standard move set (SMS) \cite{Hilhorst}
and the one involving moves of three beads of the BB
\cite{Betancourt}. The latter will be referred to as MS3.
Typical moves of SMS and MS3 are shown in Fig. 2. The SMS contains
the single moves and the crankshaft motion.
In addition to these moves, MS3 contains two and three-monomer
moves. {\em The MS3 is shown to be more efficient than  SMS} due to 
its great flexibility.
Second, 
we study {\em the effect of SCs on the scaling
of $t_f$} using cubic lattice Go models. At $T=T_{min}$
we have found $\lambda \approx 3.7$ which is higher than that
for Go models without SCs. In other words, the
scaling of folding times with $N$ for chains with and without SCs
is dramatically different. 

The energy of a conformation with SC can be written as \cite{Klimov}
\begin{eqnarray}
H \; = \;  \epsilon _{bb} \sum_{i=1,j>i+1}^{N} \, \delta _{r_{ij}^{bb},a} \,
+ \epsilon _{bs} \sum_{i=1,j\neq i}^{N} \, \delta _{r_{ij}^{bs},a} \nonumber\\
+ \epsilon _{ss} \sum_{i=1,j>i}^{N} \, \delta _{r_{ij}^{ss},a} \; ,
\label{eqn2}
\end{eqnarray}
where $\epsilon _{bb}, \epsilon _{bs}$ and $\epsilon _{ss}$ are BB-BB,
BB-SC and SC - SC interaction energies.
The quantities $r_{ij}^{bb}, r_{ij}^{bs}$ and $r_{ij}^{ss}$ are 
the spatial distance
between the $i^{th}$ and $j^{th}$ residues of the  BB-BB,
BB-SC and SC - SC, respectively.
Self-avoidance, i.e. any BB and SC beads cannot occupy the same 
lattice site more than once, is imposed.

\begin{figure}
\epsfxsize=2.8in
\centerline{\epsffile{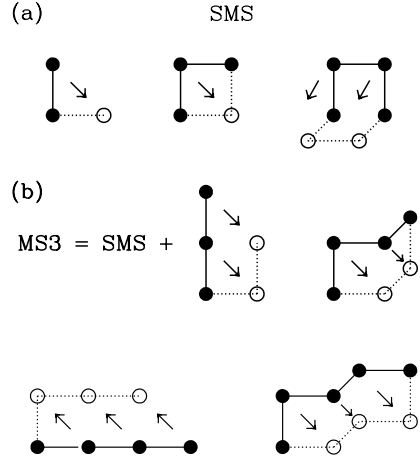}}
\caption{(a) The single monomer and the crankshaft moves in the SMS.
(b) Typical moves in the MS3. It includes all of the moves of the SMS and
the two- and three-bead moves. The open 
circles denote new positions.}
\end{figure}

To study the effect of SCs we will consider Go models \cite{Go} in 
which $\epsilon _{bb}, \epsilon _{bs}$ 
and $\epsilon _{ss}$ are chosen to be -1 for native contacts 
and 0 for non-native ones. Despite this simplification, as speculated by
Tanaka \cite{Tanaka}, the Go models capture certain generic properties
of protein folding. This is due to the fact that the geometry of the native
state, and  not details of interaction between amino acids, seems to 
play an important role in determining folding pathways and rates \cite{Baker}.

The rules for Monte Carlo backbone moves used in our kinetic
study are as 
follows. In the SMS the possible moves are tail flip (20$\%$), 
corner flip (20$\%$)
and crankshaft (60$\%$) \cite{Li}.
For MS3  we  first enumerate all possible non-overlapping
conformations of linear chains up to a maximum of $r_m$+2 residues, 
where $r_m (=3)$ is the maximum number of residues allowed to move in a single
Monte Carlo step. The number of $r$ residues
to move is selected with an exponentially decaying probability 
\cite{Betancourt}
\begin{equation}
P_r \; \; = \; \; 
\frac{(\gamma - 1)\gamma ^{r_m}}{(\gamma ^{r_m} - 1) \gamma ^r} \; ,
\end{equation}
where $r=1, \ldots ,r_m$,
$\gamma$ is set to be 1.35 which
is an optimal value for folding. Once a segment of $r$ residues
is randomly chosen, one of the $b_r$ neighboring conformations is
selected with uniform 
probability as a new local conformation.
Since $b_r$  
depends on the 
initial conformation of the segment, the move is accepted with probability 
$b_r/b_r^{max}$ in order to satistfy the detail balance condition.
$b_r$ and $b_r^{max}$ depend on whether the segment is bounded on one or 
both sides. 
If new positions for BB beads are allowed, 
then SC moves are determined. The Metropolis
criteria is applied once the moves of both BB and SC
beads are allowed geometrically. If an attempt involving a move of BB
monomers fails then one tries to move the corresponding SC beads
simutaneously \cite{Betancourt}.

We study the dependence of $t_f$ and  $t_f^{bb}$, the backbone 
folding time, on $N$. $t_f$ is defined as the median of MC times
to reach all (BB-BB, BB-SC and SC-SC)
native contacts, whereas $t_f^{bb}$ is the median of first passage times
for BB-BB native contacts.
 
Fig. 3 shows the temperature dependence of the folding times obtained
by the MS3 for
 the sequence whose native conformation is shown in Fig. 1.
Clearly, the U-shape also holds for folding times of the BB.
For the sequence studied $T_{min}$ is the same for $t_f$ and $t_f^{bb}$ but
it may not be valid for other sequences. The bottom of the $U$-shape curve
is rather wide and this is a specific feature of Go and other optimized
sequences.  At low and
high temperatures the native backbone contacts form before 
the whole chain folds.
In this paper we focus on the scaling of folding times at
$T=T_{min}\approx T_F$.

\begin{figure}
\epsfxsize=2.8in
\centerline{\epsffile{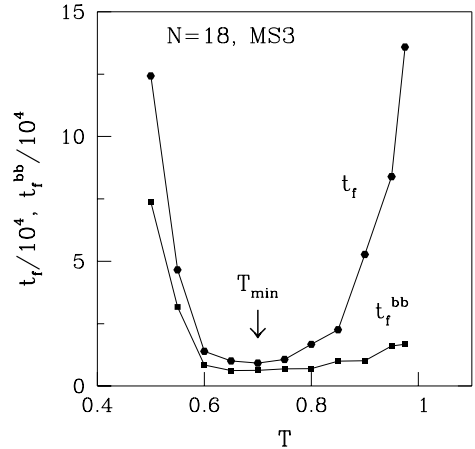}}
\caption{The temperature dependence of $t_{f}$ and $t_f^{bb}$ 
using  MS3 dynamics for 
the sequence shown in Fig. 1. 
The arrow indicates $T_{min}$.}
\end{figure}

Fig. 4 shows the dependence of folding times obtained by two types of
dynamics. To calculate the folding times we 
computed the distribution of
first passage times from one hundred invidual trajectories.
The number
of targets we used for
$N=9$, 15, 18, 24, 28, 32 and 40  are 100, 50, 50, 20, 17, 15  and 15,
 respectively.

We obtain $\lambda = 3.6 \pm 0.2$ and $3.7 \pm 0.2$
for MS3 and SMS, respectively.  The
power law behavior (\ref{eqn1}) is also valid  for BB folding,
i.e. $t_f^{bb} \sim N^{\lambda_{bb}}$. We found 
$\lambda_{bb} = 3.9 \pm 0.3$ and $3.9 \pm 0.3$
for MS3 and SMS, respectively.
Within error bars $\lambda = \lambda_{bb}$ as expected.
Interestingly, exponents $\lambda$ and $\lambda_{bb}$ obtained by
two different move sets remain the same.
Since exponent $\lambda$ for Go models with SCs
is higher than that for models without SCs
\cite{Shakhnovich,Cieplak} we conclude that
they show different folding kinetics. Such differences
apparently become much more enhanced 
for more realistic models of proteins \cite{MSLi}.
Thus, it is reasonable to expect that 
dense side chain packing may provide additional
folding barriers. 

Although the scaling exponents are almost the same for two
move sets, 
the dynamics have the visible effect on absolute values of folding times. 
This is demonstrated in the upper panel of Fig. 5 where the dependence of   
$t_f^{SMS}/t_f^{MS3}$
on $N$ is shown. The folding times
obtained by the SMS are about two times longer 
than those by the MS3 but
the real gain in CPU time is about one and half
times due to the increased complexilty of MS3.
So, the MS3 proposed by Betancourt and
Thirumalai is more efficient for folding lattice models, 
because MS3 involves more possible moves making dynamics more flexible.

\begin{figure}
\epsfxsize=2.8in
\centerline{\epsffile{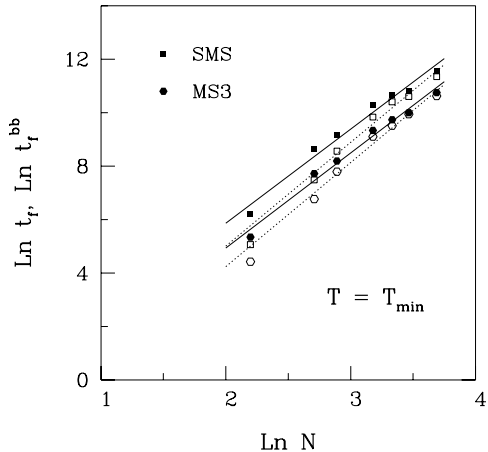}}
\caption{Scaling of $t_f$ and  $t_f^{bb}$
at $T=T_{min}$. The results were obtained by
the SMS and MS3.
The closed and open symbols
denote $t_f$ and $t_f^{bb}$.
Straight solid and dotted lines are linear fits for $t_f$ and  $t_f^{bb}$,
respectively. 
The results are averaged over  100, 50, 50, 20, 17, 15  and 15 target
conformations for $N$=9, 15, 18, 24, 28 , 32 and 40, respectively.}
\end{figure}

\begin{figure}
\epsfxsize=2.8in
\centerline{\epsffile{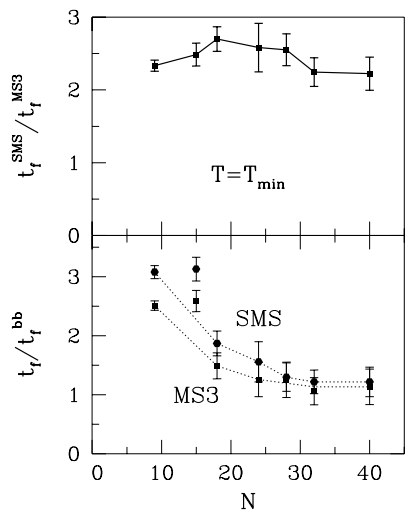}}
\caption{ The dependence of $t_f^{SMS}/t_f^{MS3}$
on N (upper panel). The statistics are the same as in Fig. 4. 
The ratio $t_f/t_f^{bb}$ obtained by the SMS and
MS3 is shown in the lower panel.The results are
obtained at $T = T_{min}$. }
\end{figure}

The relation between
time scales to fold the BB and the whole sequence 
is demonstrated in the lower panel in Fig. 5. 
The results for both move sets show that the 
BB native
contacts of short chains can be reached relatively fast. 
As $N$ increases the folding times
$t_f$ and $t_f^{bb}$ become comparable.

We now discuss the implication of our results for experiments.
Recently, Plaxco {\em et al.} \cite{Plaxco} suggested that
for real proteins there is little correlation between the folding times and
chain lengths. On the other hand, 
we have shown \cite{MSLi} that chain length dependence must 
be incorporated to improve the  correlation between folding rates and
contact order \cite{Plaxco}. 
Off-lattice protein models \cite{Cieplak1} also implicate 
a power law behavior (\ref{eqn1}) 
at $T_{min}$. Both experimental \cite{Plaxco}
and simulation \cite{Cieplak1} results are, however, based on one
value of the chain length but not on large statistics. So the question
about the scaling of $t_f$ on $N$ for real proteins remains open.

In conclusion, we have studied the scaling properties of Go sequences with
SCs by two types of dynamics. 
The MS3 has proved to be a better choice for studying folding than the SMS.
The models with
and without SCs show different kinetic properties, such  as distinct 
scalings with $N$. 

Fruitful discussions with M. Betancourt and R. Dima are gratefully 
acknowledged. MSL thanks T.X. Hoang for technical help in RASMOL plot. 
This work was supported by KBN (Grant No. 2P03B-146-18).


\vspace{0.5cm}

\begin{thebibliography}{00}
\bibitem{Dill} K. A. Dill {\em et al}, Protein Science 4 (1995) 561-602.

\bibitem{Thirumalai0} D. Thirumalai, and D. K. Klimov, 
Curr. Opin. Struc. Biol. 9 (1999) 197-207.

\bibitem{Thirumalai} D. Thirumalai, J. Phys. I (France)  5 (1995) 1457-1467.

\bibitem{Shakhnovich} A. M. Gutin, V. I. Abkevich, and E. I. Shakhnovich,
Phys. Rev. Lett. 77 (1996) 5433-5436.

\bibitem{Cieplak} M. Cieplak, T. X. Hoang and M. S. Li, Phys. Rev. Lett.  83
(1999) 1684-1687.

\bibitem{Go} N. Go and H. Abe, Biopolymers 20 (1981) 1013-1031.

\bibitem{Klimov} D. K. Klimov and D. Thirumalai, Folding and Design 3 (1998)
127-139.

\bibitem{Hilhorst} H. J. Hilhorst and J. M. Deutch, J. Chem. Phys. 63 (1975)
5153-5161. 

\bibitem{Betancourt} M. R. Betancourt, J. Chem. 109 (1998) 1545-1554;
M. R. Betancourt and D. Thirumalai, preprint.

\bibitem{Tanaka} S. Tanaka, Proc. Natl. Acad. Sci. 96 (1999) 11698-11700.

\bibitem{Baker} D. Baker, Nature 405 (2000) 39-42.

\bibitem{Li} L. Li, L. Mirny and E. I. Shakhnovich, Nature Struct. Biol.
7 (2000) 336-344.

\bibitem{MSLi} M. S. Li, D. K. Klimov, and D. Thirumalai, preprint

\bibitem{Plaxco} K. W. Plaxco, K. T. Simons, and D. Baker, J. Mol. Biol.
277 (1998) 985-994.

\bibitem{Cieplak1} M. Cieplak and T. X. Hoang, Proteins 44 (2001) 20-25.

\end{thebibliography}
\end{document}